\documentclass[12pt,nofootinbib,aps,amsmath,amssymb,preprint,showpacs]{revtex4}

\usepackage[T1]{fontenc} \usepackage[latin1]{inputenc}
\usepackage{amsmath,color,ulem} \usepackage{graphicx}
\usepackage{epsf,bm} \usepackage{amssymb}
\newcommand\underrel[2]{\mathrel{\mathop{#2}\limits_{#1}}}
\usepackage{tabulary}
\newcolumntype{K}[1]{>{\centering\arraybackslash}p{#1}}
\usepackage{multirow}
\begin{document}

\title{Critical Dynamical Exponent of the Two-Dimensional
Scalar $\phi^4$-Model with Local Moves}
\author{Wei Zhong$^\dagger$} 
\email{w.zhong1@uu.nl}
\author{Gerard T. Barkema$^\dagger$}
\author{Debabrata Panja$^\dagger$}
\author{Robin C. Ball$^\ddagger$}
 
\affiliation{ $^\dagger$Department of Information and Computing
  Sciences, Utrecht University, Princetonplein 5, 3584 CC Utrecht, The
  Netherlands\\ 
  $^\ddagger$Department of Physics, University of Warwick, Coventry CV4 7AL, United Kingdom}

\date{\today} 

\begin{abstract}
We study the scalar one-component two-dimensional (2D) $\phi^4$-model
by computer simulations, with local Metropolis moves. The equilibrium
exponents of this model are well-established, e.g. for the 2D
$\phi^4$-model $\gamma= 1.75$ and $\nu= 1$. The model has also been
conjectured to belong to the Ising universality class. However, the
value of the critical dynamical exponent $z_c$ is not settled. In this
paper, we obtain $z_c$ for the 2D $\phi^4$-model using two independent
methods: (a) by calculating the relative terminal exponential decay
time $\tau$ for the correlation function $\langle
\Phi(t)\Phi(0)\rangle$, and thereafter fitting the data as $\tau \sim
L^{z_c}$, where $L$ is the system size, and (b) by measuring the
anomalous diffusion exponent for the order parameter, viz., the
mean-square displacement (MSD) $\langle \Delta \Phi^2(t)\rangle\sim
t^c$ as $c=\gamma/(\nu z_c)$, and from the numerically obtained value
$c\approx 0.80$, we calculate $z_c$. For different values of the
coupling constant $\lambda$, we report that $z_c=2.17\pm0.03$ and
$z_c=2.19\pm0.03$ for the two methods respectively. Our results
indicate that $z_c$ is independent of $\lambda$, and is likely
identical to that for the 2D Ising model. Additionally, we demonstrate
that the Generalized Langevin Equation (GLE) formulation with a memory
kernel, identical to those applicable for the Ising model and
polymeric systems, consistently captures the observed anomalous
diffusion behaviour.
 
\end{abstract}

\maketitle
 
\section{Introduction \label{sce1}}
The $\phi^4$-model has become one of the most useful tools in studying of critical
phenomena \cite{kleinert,kaupuzes1,pelissetto,amit}. In two
dimensions, the lattice version of the $\phi^4$-model is defined by
the action ${\cal S}$ and Hamiltonian ${\cal H}$ as
\begin{equation}
{\cal S}=\frac{{\cal H}}{k_B T}=-\beta\sum_{\langle ij\rangle}\phi_i\phi_j+\sum_i[\phi_i^2+\lambda(\phi_i^2-1)^2],
  \label{a1} 
\end{equation} 
where $-\infty<\phi_i<\infty$ is the dynamical variable at site $i$, $\beta$ and $\lambda$ are two model constants. The
summation of the first term in the r.h.s of Eq. (\ref{a1}) runs over
all the nearest\text{-}neighbour spins, and for an $L\times L$ square
lattice $0\leq(i,j)< L$.
The order parameter for the $\phi^4$-model is defined as $\Phi=\sum_i \phi_i$, and the dynamics of the model is given by \cite{zinn,lin}
 \begin{eqnarray}
\dot{\phi_i} &=& - \Omega \frac{\partial {\cal S}}{\partial \phi_i}+\xi(i,t)\label{dyn}\\
\langle \xi(i,t) \xi(i',t')\rangle&=&2\Omega \delta(i-i')\delta(t-t'),  
\label{dyn1}  
\end{eqnarray}
where $\xi(i,t)$ is a Gaussian noise term, and $\Omega$ represents the dissipation constant, which is related to the noise term by the Fluctuation-Dissipation relation (\ref{dyn1}).

The equilibrium properties of the model in relation to the critical
phenomenon are well\text{-}studied. Earlier investigations of the two
dimensional (2D) and three dimensional (3D) lattice $\phi^4$-model
have indicated that the critical exponents $\gamma$ and
$\nu$ are the same as these for the Ising model, e.g. in 2D, $\gamma =
1.75$ and $\nu = 1$ \cite{michev,hasenbusch,kaupuzes2}. Simultaneously,
Monte Carlo simulations of the 2D lattice $\phi^4$-model have
supported the idea that the $\phi^4$-model belongs to the Ising
universality class \cite{mehling}. Despite these advances in the
equilibrium properties of the model, its critical dynamical properties
are not settled.
 
As for the critical dynamical exponent, Bl\"{o}te and Nightingale
\cite{blote1} have analyzed three variations of Ising\text{-}type
models with next\text{-}nearest\text{-}neighbour interactions, and
found that they share the same critical exponents, not only $\gamma$
and $\nu$, but also the critical dynamical exponent $z_c$. Further
works have supported their results both in 2D and 3D
\cite{blote2,hasenbusch2,hohenberg}. For the 2D Ising model $z_c$ has
been determined quite precisely as $z_c=2.1665\pm 0.0012$
\cite{blote3}. For the critical dynamical exponent of the 2D $\phi^4$-model, $z\approx 2$ was mentioned in Ref. \cite{wolff}, and the
$\epsilon$\text{-}expansion method has shown that $z_c\in (2.04,2.14)$
\cite{nalimov}. Further, $z_c$ has been measured using the heat bath
algorithm, yielding $z_c=1.9\pm 0.21$ \cite{ brower}. In short, the
value of the critical dynamical exponent for the $\phi^4$-model still
remains to be determined with higher precision.
 
In this paper, we study the one-component 2D scalar $\phi^4$-model by
computer simulations, i.e., Eq. (\ref{a1}), with local Metropolis
moves. In order to settle the value of $z_c$, we employ two
independent methods: (a) we calculate the relative terminal
exponential decay time $\tau$ for the correlation function $\langle
\Phi(t)\Phi(0)\rangle$, and thereafter fit the data as $\tau \sim
L^{z_c}$, where $L$ is the system size, (b) we measure the mean-square
displacement (MSD) of the order parameter $\langle \Delta
\Phi^2(t)\rangle\sim t^c$ with $c=\gamma/(\nu z_c)$, and from the
numerically obtained value $c\approx 0.80$ we calculate $z_c$. We report
that $z_c=2.17\pm0.03$ and $z_c=2.19\pm0.03$ for the two methods
respectively. Our results suggest that $z_c$ is independent of
$\lambda$, and is likely identical to that for the 2D Ising model.

Further, the numerical result $\langle \Delta \Phi^2(t)\rangle\sim
t^{0.80}$ at the critical point means that $\Phi(t)$ undergoes
anomalous diffusion. We argue that the physics of anomalous diffusion
in the $\phi^4$-model at critical point is the same as for polymeric
systems and the Ising model \cite{panja3,panja4,panja2a,zhong}, and
therefore a GLE formulation that holds for the Ising model at
criticality and for polymeric systems must also hold for the
$\phi^4$-model. We obtain the force autocorrelation function for the
$\phi^4$-model at $\dot\Phi=0$, and the results allow us to
demonstrate the consistency between anomalous diffusion and its GLE
formulation.
 
The paper is organised as follows. In Sec. \ref{sec2} we introduce the
$\phi^4$-model and the dynamics, and then show the results of the
correlation term $\langle\Phi(t)\Phi(0)\rangle$ and the mean-square
displacement of the order parameter; from both we measure the critical
dynamical exponent. In Sec. \ref{sec3} we briefly explain how the
restoring force works, which naturally leads us to the Generalised
Langevin Equation (GLE) formulation for the anomalous diffusion in the
$\phi^4$-model, and verify the GLE formulation for anomalous
diffusion. The paper is concluded in Sec. \ref{sec4}.

\section{The Measurement of the Critical Dynamical Exponent \label{sec2}} 

\subsection{The Model and the Dynamics \label{sec2a}}

We consider the scalar one-component two-dimensional $\phi^4$-model on
an $L\times L$ square lattice with periodic boundary conditions. The
action is introduced in Eq. (\ref{a1}), and in this paper we focus on
$\lambda\leq 1$.

We simulate the dynamics of the system, i.e. Eq. (\ref{dyn}), using Monte Carlo moves, with the Metropolis algorithm: we randomly select a site $i$, for which we try to change
the existing value $\phi_i$ to a new value $\phi_i'$,
given by
\begin{equation}
 \phi_i'=\phi_i+\Delta\phi\left(r-\frac{1}{2}\right),
 \label{b1}
\end{equation}
where $r$ is a random number uniformly distributed within [0,1), and
  following Refs. \cite{hasenbusch,kaupuzes2}, we set $\Delta
  \phi=3$. The resulting change of the action $\Delta {\cal S}$ after
  every attempted change in $\phi_i$ is calculated. The
  move is accepted if $\Delta {\cal S}\le0$; if not, then the move is
  accepted with the usual Metropolis probability $e^{-\Delta {\cal
      S}}$. With $\Delta \phi=3$, the acceptance rates are between
  $40\%$ and $60\%$.

In this paper, all simulations have been performed on a $3.40$GHz desktop PC running Linux. We mainly focus on three
different values of $\lambda$, i.e. $\lambda=0.1,0.5, 1.0$. The corresponding
critical coupling constant $\beta_c$, obtained in
Refs. \cite{kaupuzes2,bosetti}, are listed in table \ref{tab1}.

\begin{table}[h] 
\begin{tabular}{K{3.5cm}|K{3.5cm}} 
\hline\hline
$\lambda$ & Value of $\beta_c$ \\
\hline\hline
0.1& 0.60647915(35)\\
\hline
0.5& 0.686938(10) \\
\hline 
1.0& 0.680601(11) \\
\hline

\end{tabular} 
\caption{The value of $\beta_c$ for $\lambda=0.1,0.5$ and $1.0$
  \cite{kaupuzes2,bosetti}. \label{tab1}}
\end{table}

Next, we use two independent methods to measure the dynamical exponent $z_c$.

\subsection{Measurement of the Correlation function $\langle\Phi(t)\Phi(0)\rangle$ \label{sec2b}}

In the first method, we measure the correlation function $\langle
\Phi(t)\Phi(0)\rangle$ of the order parameter. To obtain the
corresponding data, we run our simulations for $5\times 10^7$ Monte Carlo steps per lattice site to thermalise the system. Subsequently, we
keep taking snapshots of the system at regular intervals over a total time of $5\times 10^8$ Monte Carlo steps per lattice site, and compute
the order parameter $\Phi$ at every snapshot. From this data set we calculate
$\langle\Phi(t)\Phi(0)\rangle$. 

 We use system sizes $L=30,40,...,90$ for each value of $\lambda$. The required CPU time is about $45$ minutes for $L=30$, reaching about $6$ hours for $L=90$. 

\begin{figure*}
\includegraphics[width=0.43\linewidth]{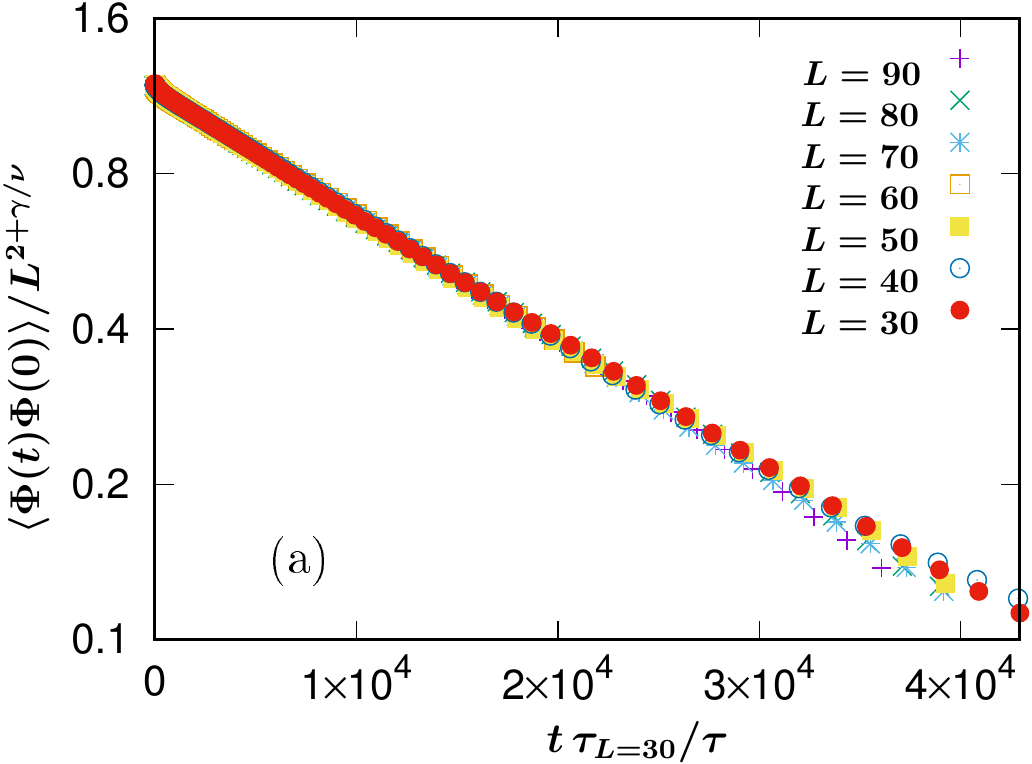}
  \hspace{5mm}
  \includegraphics[width=0.43\linewidth]{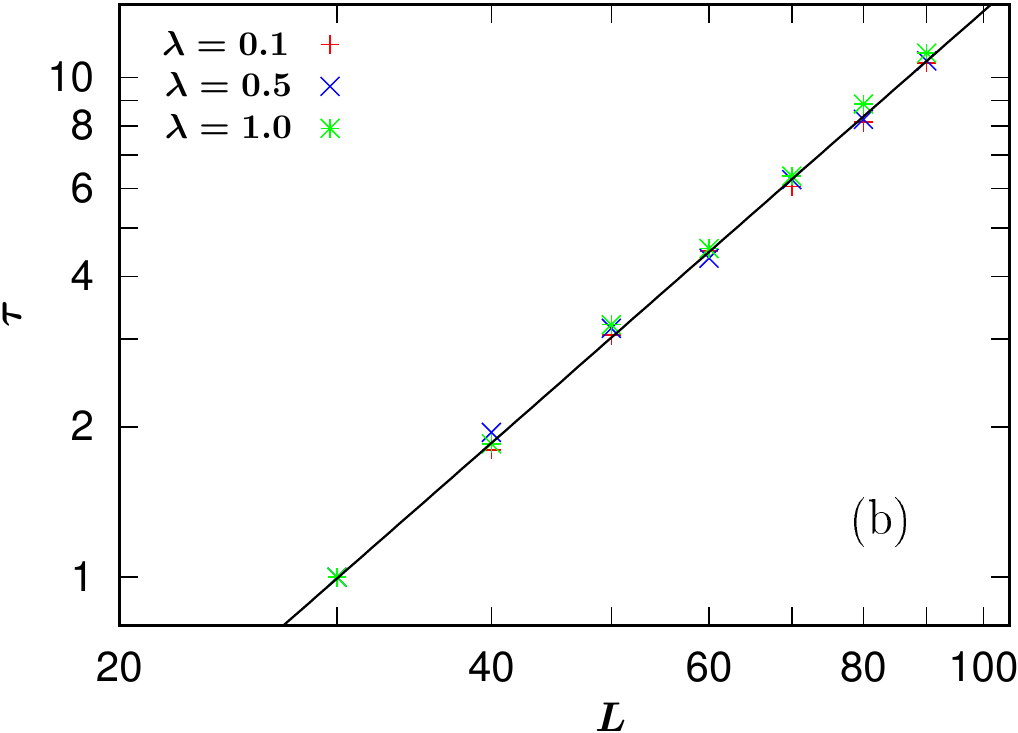}
  \caption{(color online) Measurement of $z_c$ from the $Q(t)$ data
    for different values of $\lambda$ at the critical point. (a) An
    example of the measurement process to obtain the relative value of
    terminal decay time $\tau$. In this figure $\lambda=0.1$ and
    $\tau_{L=30}$ is the terminal decay time of reference system size
    $L=30$. The data collapse (to the reference) is obtained by
    adjusting the relative values of terminal decay time for other
    system sizes; (b) Measurement of $z_c$ by fitting the data of the
    relative value of decay time $\tau$ as $\tau \sim L^{z_c}$.The
    symbols represent the simulation results of $\tau$ and the solid
    line corresponds to $\tau\sim L^{2.17}$.}
  \label{corrleng}
\end{figure*}

At long times we expect $\langle \Phi(t)\Phi(0)\rangle$ to behave as
$\langle \Phi(t)\Phi(0)\rangle/\langle \Phi(0)\Phi(0)\rangle\sim
\exp(-t/\tau)$, and define $Q(t)=-\ln\,[\langle
  \Phi(t)\Phi(0)\rangle/\langle \Phi(0)\Phi(0)\rangle]$, leading us to
expect
\begin{equation}
Q(t)\sim t/\tau.
\label{b3}
\end{equation}
We then calculate the relative value of terminal decay time $\tau$
by collapsing the $Q(t)$ data to a reference for every value of
$\lambda$. More explicitly, for every value of $\lambda$ we choose the
$Q(t)$ data for $L=30$ as reference, set its $\tau$-value to unity,
and then collapse the rest of the $Q(t)$ for other values of $L$ to
that reference, which yields us the relative value of $\tau$ for that
value of $\lambda$. As an example, Fig. {\ref{corrleng}(a)}
demonstrates this procedure: with a properly chosen relative value of
$\tau$, the $\langle\Phi(t)\Phi(0)\rangle$ data for different system
sizes collapse to the data of $L=30$.

At the critical temperature $\tau\sim \xi^{z_c}$, where $\xi$ is the
correlation length. According to finite-size scaling theory, for
finite system sizes $\xi$ needs to be replaced by $L$, i.e.,
\begin{equation}
\tau\sim L^{z_c}.
\label{b4}
\end{equation}
\begin{table}[h] 
\begin{tabular}{K{2.5cm}|K{2.5cm}} 
\hline\hline
$\lambda$ & $z_c$  \\ 
\hline\hline 
$0.1$ & 2.17 $\pm$ 0.03\\
\hline 
$0.5$ & 2.15 $\pm$ 0.03   \\
\hline 
$1.0$ & 2.20 $\pm$ 0.03  \\
\hline 
\end{tabular}
\caption{The measured values of $z_c$ for the 2D $\phi^4$-model at
  different $\lambda$. The critical dynamical exponent $z_c$ is
  calculated by fitting the data of the relative value of $\tau$ as
  $\tau\sim L^{z_c}$. The results indicate that the value of $z_c$ is
  likely independent of $\lambda$, which allows us to produce a single
  estimate of $z_c$, viz., $z_c=2.17\pm0.03$ (see main
  text). \label{tab2}}
\end{table} 

The critical dynamical exponent $z_c$ is calculated by fitting the
data of the relative value of $\tau$ with Eq. (\ref{b4}). Results of
this procedure are shown in Fig. \ref{corrleng}(b). The corresponding
values of $z_c$ can be found in Table. \ref{tab2}. The error bars in
Table \ref{tab2} are obtained from the best fits of
Fig. \ref{corrleng}(b). These results indicate that the value of $z_c$
is likely independent of $\lambda$. If we do assume that, then we can
combine the different numerical values for different $\lambda$ to
produce a single estimate of $z_c$, viz., $z_c=2.17\pm0.03$.

\subsection{Mean-Square Displacement of the Order Parameter  \label{sec2c}}

In the second method, we focus on the measurement of the
mean\text{-}square displacement of the order parameter at time $t$,
given by
\begin{equation}
\langle\Delta \Phi^2(t)\rangle = \langle [\Phi(t)-\Phi(0)]^2\rangle.
\label{b5}
\end{equation}
To obtain the data of the MSD of the order parameter, we first
thermalise the system with $2\times 10^8$ Monte Carlo moves per lattice site, then measure $\langle\Delta \Phi^2(t)\rangle$
in a further simulation over $2\times 10^9$ Monte Carlo moves per lattice site, using the shifting time window method. 

 For each value of $\lambda$, three different system sizes are used: $L=40, 80, 160$. For $L=40$, the simulation runs for about $5$ hours, and it takes about 3 days to obtain the results for $L=160$.

At short times ($t\approx 1$), the individual changes of $\Phi$ are
uncorrelated; i.e., the mean-square displacement (MSD) of the order
parameter must behave as $\langle\Delta \Phi^2(t)\rangle \sim L^dt$,
where $d=2$ is the spatial dimension of the system.

At long times, $t\gtrsim L^{z_c}$, we expect
$\langle\Phi(t)\Phi(0)\rangle=0$, which means that
\begin{equation}
\langle\Delta \Phi^2(t)\rangle\underrel{t\gg L^{z_c}}\approx 2\langle
\Phi(t)^2\rangle \sim L^{d+\gamma/\nu},
\label{b6}
\end{equation} 
which is an equilibrium quantity.

If we assume that the MSD is given by a simple power-law in the
intermediate time regime ($1\lesssim t\lesssim L^{z_c}$), then we have
\begin{equation}
\langle\Delta \Phi^2(t)\rangle\sim t^c,
\label{b7}
\end{equation}
where $c=\gamma/(\nu z_c)$. Note that exactly the same behavior has
been found in the Ising model \cite{walter,zhong}.

In order to measure the value of the exponent $c$ from $\langle \Delta
\Phi^2(t)\rangle$, we need to focus on the intermediate time regime,
i.e. we consider the MSD data in $(t_{min},t_{max})$ to estimate the
exponent. From these data we calculate the exponent $c$ as numerical
derivative as $\displaystyle c=\frac{1}{t_{max}-t_{min}}\sum\limits_{
  t=t_{min}}^{t_{max}-1}\frac{ \ln\,\langle\Delta
  \Phi^2(t+1)\rangle-\ln\,\langle\Delta
  \Phi^2(t)\rangle}{\ln\,(t+1)-\ln\,t}$. In order to estimate $z_c$
for different $\lambda$ from these data, we use the data from the
largest system size so that we can limit the influence of finite-size
effects. From the numerically obtained $c$ we calculate $z_c$ and
$c=\gamma/(\nu z_c)$, which we present in Table. \ref{tab3}. These
results, too, indicate that the value of $z_c$ is likely independent
of $\lambda$. If we do assume that, then we can combine the different
numerical values for different $\lambda$ to produce a single estimate
of $z_c$, viz., $z_c=2.19\pm0.03$.
\begin{table*}
\begin{tabular}{K{2.5cm}|K{2.5cm}} 
\hline\hline
$\lambda$ & $z_c$  \\ 
\hline\hline 
$0.1$ & 2.20$\pm$0.03\\
\hline 
$0.5$ & 2.18$\pm$0.02   \\
\hline 
$1.0$ & 2.20$\pm$ 0.04  \\
\hline 
\end{tabular}
\caption{The critical dynamical exponent $z_c$, which is obtained from
  the numerically obtained $c$ with $c=\gamma/(\nu z_c)$, for the 2D
  $\phi^4$-model at different $\lambda$. The results, too, indicate
  that the value of $z_c$ is likely independent of $\lambda$, which
  allows us to produce a single estimate of $z_c$, viz.,
  $z_c=2.19\pm0.03$ (see main text). \label{tab3}}
\end{table*}
\begin{figure*}
\includegraphics[width=0.43\linewidth]{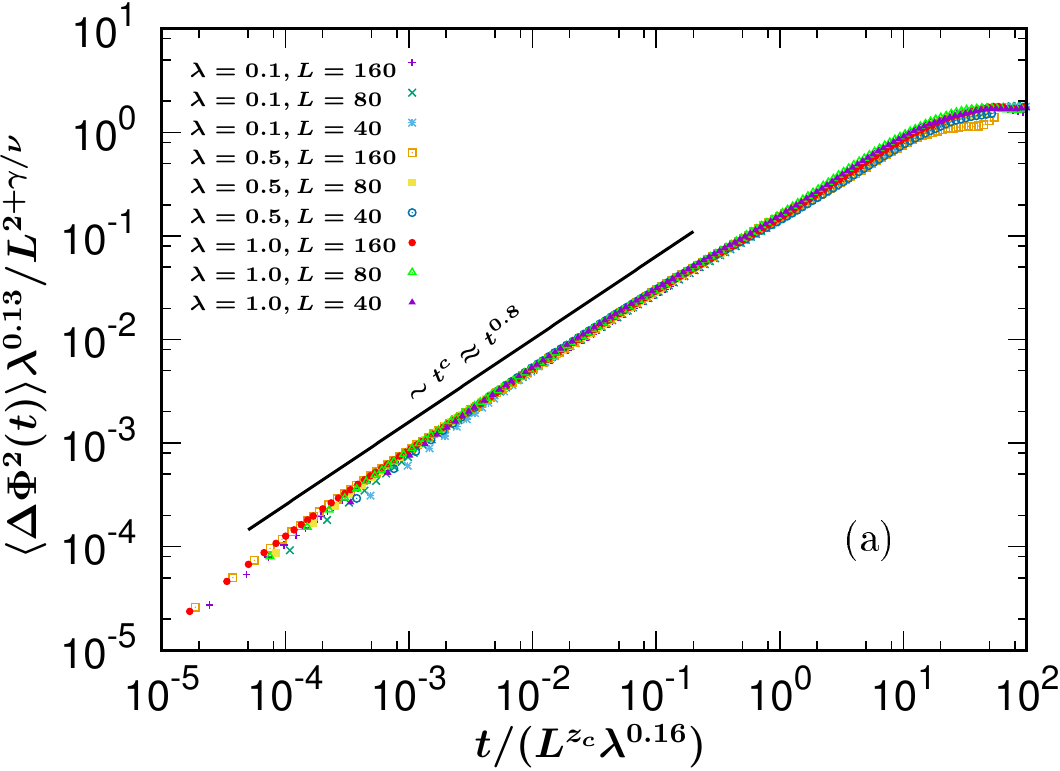}
\hspace{5mm}
\includegraphics[width=0.4 \linewidth]{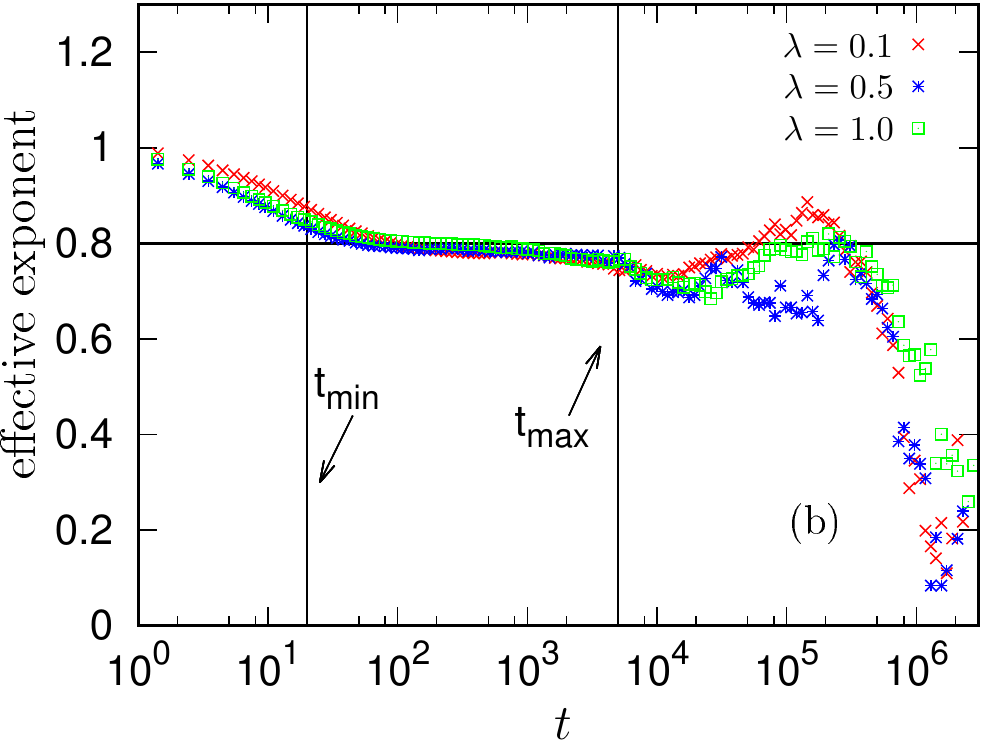}
\caption{(color online) (a) The mean-square displacement (MSD) of the
  order parameter $\langle \Delta \Phi^2(t)\rangle$ in the $\phi^4$-model at the critical point. The MSD scales as $\langle \Delta
  \Phi^2(t)\rangle\sim L^2t^c$ where $c=\gamma/(\nu z_c)$. The $x$ and
  $y$ axes are scaled with $\lambda$-dependent numerical scale factors
  to achieve good quality data collapse for different $\lambda$. The
  solid line denotes the power-law behavior shown in Eq. (\ref{b6}).
  (b) The effective exponents, i.e., the numerically differentiated
  $d\,\ln(\langle\Delta\Phi^2(t)\rangle) / d\,\ln(t)$, of the MSD of
  the order parameter for different $\lambda$. The system sizes are
  $L=160$ and the time period for calculating the exponent $c$ is
  $t\in(t_{min},t_{max})$, where $t_{min}\sim 1$ and $t_{max}\sim
  L^{z_c}$.  } \label{msd}
\end{figure*}

The corresponding data for the MSD of $\Phi(t)$ for $80\leq L \leq
160$ for different values of $\lambda$ are shown in
Fig. \ref{msd}. The small deviation in Fig. \ref{msd} at late times is
caused by periodic boundary conditions: they are different when free
boundary conditions are utilised. (Exactly the same effect has been
observed in our earlier work on the Ising model
\cite{zhong}. Verification of the boundary effects is therefore not
shown here, since the deviations from the power-law do not scale with
$L$, and consequently are not relevant in the scaling limit.)

In conclusion, the critical dynamical exponent $z_c$ obtained with two
independent methods demonstrate that $z_c = 2.17\pm0.03$ or $z_c =
2.19\pm0.03$ for different values of $\lambda$ in the 2D scalar
$\phi^4$-model. Both results are consistent to the value of $z_c$ for
the 2D Ising model ($2.1665\pm0.0012$). In other words, our results
indicate that $z_c$ is independent of $\lambda$, and is likely
identical to that for the 2D Ising model.
 
\section{The GLE formulation of the anomalous diffusion in the $\phi^4$-model \label{sec3}}

In Sec. \ref{sec2c} we numerically obtained that, in the intermediate
time regime, the MSD of the order parameter in the $\phi^4$-model
behaves as
\begin{equation}
\langle\Delta \Phi^2(t)\rangle \sim L^2t^{0.80}.
\label{c1}
\end{equation}
This means that, at the critical point, the order parameter exhibits
anomalous diffusion. The same behavior has been observed in the Ising
model \cite{walter}. The physics of anomalous diffusion in the Ising
model has been thoroughly analysed in Ref. \cite{zhong}, where it has
also been demonstrated that the physics is identical to that for
polymeric systems
\cite{panja2a,panja3,panja4,panja1,panja3b,panja3c,panja3d,panja2}.

Both in the Ising model and polymeric systems, the anomalous diffusion
stems from time-dependent restoring forces which lead to the GLE
formulation. Translated to the $\phi^4$-model,  the physics of
the restoring force can be described as follows.

Imagine that the order parameter locally changes by an amount $\delta
\phi$ due to thermal fluctuations at $t=0$. Due to the interactions
among the spins dictated by the Hamiltonian, the system will react to
the change in $\delta \phi$. This reaction will be manifest in the two
following ways: $(a)$ the system will to some extent adjust to the
change of $\delta \phi$, however it will take some time, and $(b)$
during this time the order parameter will also readjust to the
persisting value of $\Phi$, undoing at least part of $\delta \phi$. It
is the latter that we interpret as the result of inertia that resists
change in $\Phi$, and the resistance itself acts as the restoring
force to the changes in the order parameter.
 
\subsection{The GLE formulation for the anomalous diffusion in the $\phi^4$-model \label{sec3a}}

In the Ising model and polymeric systems, the restoring force has led
to the GLE description for the anomalous diffusion
\cite{zhong,panja3,panja4}. We now import that for the $\phi^4$-model,
with a time\text{-}dependent memory function $\mu(t)$ arising out of
the restoring forces. The GLE formulation for the anomalous diffusion
is described as
\begin{subequations}
\begin{equation}
\zeta \dot{\Phi}(t) = f(t)+q_1(t)
 \label{c2_1}
\end{equation}   
\begin{equation}
f(t) = -\int_0^tdt'\mu(t-t')\,\dot{\Phi}(t')+q_2(t).
 \label{c2_2}
\end{equation}
 \label{c2}
\end{subequations}
Here $f(t)$ is the internal force, $\zeta$ is the ``viscous drag'' on
$\Phi(t)$, $\mu(t-t')$ is the memory kernel, $q_1(t)$ and $q_2(t)$ are
two noise terms satisfying $\langle q_1(t)\rangle=\langle
q_2(t)\rangle=0$, and the fluctuation-dissipation theorems (FDTs)
$\langle q_1(t)\,q_1(t')\rangle\propto\zeta\delta(t-t')$ and $\langle
q_2(t)\,q_2(t')\rangle\propto\mu(t-t')$ respectively.

Equation (\ref{c2_2}) can be inverted to write as
\begin{equation}
\dot \Phi(t)=-\int_0^t dt'\,a(t-t') f(t')+\omega(t).
\label{c3}
\end{equation}
The noise term $\omega(t)$ similarly satisfies
$\langle\omega(t)\rangle=0$, and the FDT $\langle\omega(t)
\omega(t')\rangle=a(|t-t'|)$. Then $a(t)$ and $\mu(t)$ are related to
each other in the Laplace space as $\tilde a(s)\tilde\mu(s)=1$.
 
To combine Eq. (\ref{c2_1}) and (\ref{c2_2}), we obtain
\begin{equation}
\zeta \dot{\Phi}(t)=-\int_0^tdt'\mu(t-t')\,\dot{\Phi}(t')+q_1(t)+q_2(t).
\label{c4}
\end{equation}  
or
\begin{equation}
\dot{\Phi}(t)=-\int_0^tdt'\theta(t-t')\,[q_1(t)+q_2(t)].
\label{c5}
\end{equation} 
where in the Laplace space $\tilde
\theta(s)[\zeta+\tilde\mu(s)]=1$. With $t > t'$, without any loss of
generality, using Eq. (\ref{c5}) the result of the velocity
autocorrelation is
\begin{equation}
\langle \dot{\Phi}(t) \dot{\Phi}(0)\rangle\sim \theta(t-t'),
\label{c6}
\end{equation}
 where $\theta(t)$ can be calculated by Laplace inverting the relation
 $\tilde \theta(s)[\zeta+\tilde\mu(s)]=1$.
 
If $\mu(t)$ behaves as a power\text{-}law in time with an exponential
cutoff such as
 \begin{equation}
  \mu(t)\sim L^{-2}t^{-c}\exp(-t/\tau),
  \label{c6a}
 \end{equation}
 then we have \cite{panja4}
\begin{equation}
\langle \dot \Phi(t) \dot \Phi(t') \rangle=-\theta(t-t')\sim -L^2 (t-t')^{c-2} 
\qquad \text{for}\quad t\leq \tau .
\label{c7}
\end{equation}
 By integrating Eq. (\ref{c7}) twice in time (the Green-Kubo relation), we obtain
\begin{equation}
\langle\Delta \Phi^2(t)\rangle\sim L^2 t^c
\qquad \text{for}\quad t\leq \tau .
\label{c8}
\end{equation}

The form $\mu(t)\sim L^{-2}t^{-c}$ not only obtains the anomalous
exponent for the mean-square displacement, but also the correct
$L$-dependent prefactor to achieve the data collapse in
Fig. \ref{msd}, i.e., if $\mu(t)\sim L^{-2}t^{-c}$, then
$\langle\Delta \Phi^2(t)\rangle\sim L^2 t^c$.

\subsection{Verification of the first equation of the GLE and the power\text{-}law behaviour of $\mu(t)$ \label{sec3b} }

We now numerically verify our proposed GLE formulation, including the
form of $\mu(t)$ as stated in Eq. (\ref{c6}) for anomalous diffusion
in the $\phi^4$-model.
\begin{figure}[h]
\includegraphics[width=0.43\linewidth]{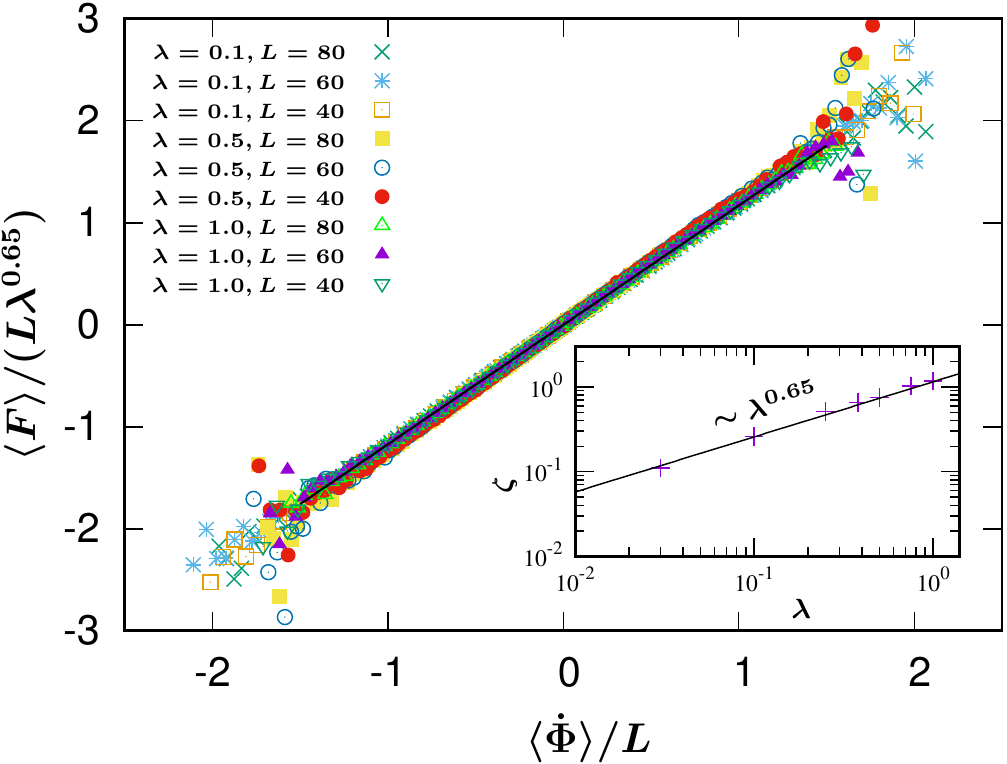}
  \caption{(color online) The linear relation Eq. (\ref{c2_1}) between
    the average internal force $\langle f\rangle$ and
    $\langle\dot{\Phi}\rangle$ for different $\lambda$. Numerically,
    we find (inset) that the viscous drag $\zeta$ behaves as
    $\zeta\sim \lambda^{0.65}$. }
  \label{firsteq}
\end{figure}

First, in order to verify Eq. (\ref{c2_1}), note that in the $\phi^4$-model, the force within the system can be directly calculated as
\begin{equation}
f=-\frac{1}{L^2}\sum_{i=0}^{N} \frac{\partial {\cal S}}{\partial
  \phi_i}\frac{\partial \phi_i}{\partial
  \Phi}=-\frac{1}{L^2}\sum_{i=0}^{N} \frac{\partial {\cal S}}{\partial
  \phi_i}.
\label{c9}
\end{equation}
By taking ensemble averages on both sides of Eq. (\ref{c2_1}) we obtain
\begin{equation}
\langle f(t)\rangle=\zeta\langle\dot{\Phi}\rangle.
\label{c10}
\end{equation}
This linear relation is demonstrated in
Fig. \ref{firsteq}. Additionally, in the inset we plot the viscous
drag $\zeta$ as a function of $\lambda$, and numerically obtain
$\zeta\sim \lambda^{0.65}$.

Next we verify the power\text{-}law behaviour of $\mu(t)$
(Eq. ({\ref{c6}})) following the FDT $\langle f(t)f(t')\rangle|_{\dot
  \Phi=0} =\mu(t-t')$.
 
We start with a thermalised system at $t=0$. For $t>0$ we fix the
value of $\Phi$ (without freezing the whole system),
which we achieve by performing non-local spin-exchange moves, i.e., at
each move, we choose two lattice site $i$ and $j$ at random, and
attempt to change the spin values to $\phi_i'=\phi_i+\Delta \phi$ and
$\phi_j'=\phi_j-\Delta \phi$. We calculate the change in the energy
$\Delta {\cal S}$ before and after every attempted move, and accept or
reject the move with the Metropolis acceptance probability. While
performing spin-exchange dynamics, we keep taking snapshots of the
system at regular intervals, and compute, at every snapshot (denoted
by $t$), the force $f(t)$ from Eq. (\ref{c9}).
\begin{figure}[h]
  \includegraphics[width=0.43\linewidth]{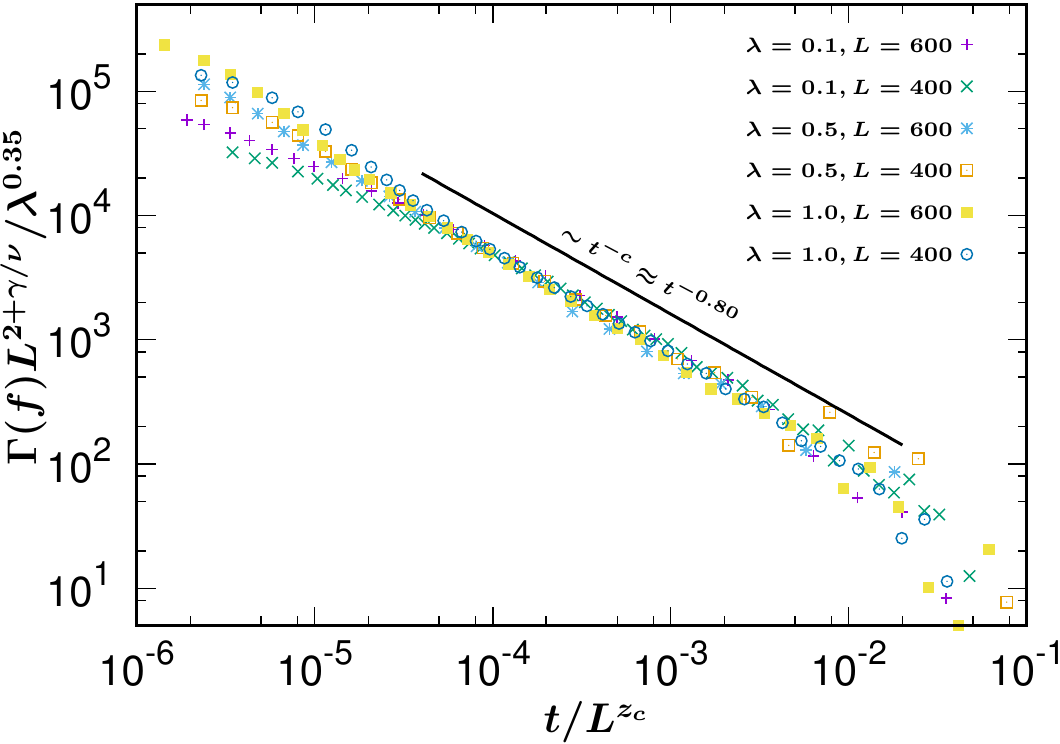}
  \caption{(color online) Behaviour $\Gamma(f)\sim t^{-0.8}$ for different
    $\lambda$ in the intermediate time regime following
    Eq. (\ref{c12}), then following the FDT, we have $\mu(t)\sim t^{-0.8}$. The extra $\lambda$-dependent factor
    $\lambda^{0.35}$ is introduced numerically to collapse the data
    for different $\lambda$ at intermediate times. Further, $z_c=2.17$
    has been used here to collapse the data.}
  \label{forceauto}
\end{figure}

We notice that since simulations are performed for finite systems with
$\Phi$ fixed at its $t=0$ value, we will in any particular run have a
non-zero value of $\langle f(t)\rangle$ acting to sustain the initial
value of $\Phi$ \cite{zhong}. Thus we calculate the quantity
\begin{equation}
\Gamma(f)=\langle f(t)f(t')\rangle-\langle f(t)\rangle \langle f(t')\rangle,
\label{c11}
\end{equation}
which we expect to represent $\mu(t-t')$ for all values of $\lambda$, i.e., 
\begin{equation}
\Gamma(f)\sim L^{-2}t^{-c} \approx L^{-2}t^{-0.80}.
\label{c12}
\end{equation}

The relation (\ref{c12}) is verified in Fig. \ref{forceauto}.

\section{Conclusion \label{sec4}}

In this paper, we have measured the critical dynamical exponent $z_c$
in the $\phi^4$-model using two independent methods: (a) by
calculating the relative terminal exponential decay time $\tau$ for
the correlation function $\langle \Phi(t)\Phi(0)\rangle$, and
thereafter fitting the data as $\tau \sim L^{z_c}$, and (b) by
measuring the mean-square displacement (MSD) of the order parameter
$\langle \Delta \Phi^2(t)\rangle\sim t^c$ with $c=\gamma/(\nu z_c)$,
and $z_c$ is calculated from the numerically obtained value $c\approx
0.80$. For different values of the coupling constant $\lambda$, we
report that $z_c=2.17\pm0.03$ and $z_c=2.19\pm0.03$ for these two
methods respectively. Our results indicate that $z_c$ is independent
of $\lambda$, and is likely identical to that for the 2D Ising model.

Further, the numerical result $\langle \Delta \Phi^2(t)\rangle\sim
t^{0.80}$ at the critical point means that $\Phi(t)$ undergoes
anomalous diffusion. We have argued that the physics of anomalous
diffusion in the $\phi^4$-model at the critical point is the same as for
polymeric systems and the Ising model \cite{panja3,panja4,zhong}, and
therefore a GLE formulation that holds for the Ising model at
criticality and for polymeric systems must also hold for the $\phi^4$-model. We obtain the force autocorrelation function for the $\phi^4$-model at $\dot\Phi=0$, and the results allow us to demonstrate the
consistency between anomalous diffusion and its GLE formulation. In
comparison to the Ising model, since $\Phi$ is a continuous order
parameter and there is a proper definition of the internal force, we
believe that the $\phi^4$-model is a better choice to verify the FDT
for the GLE formulation.

Finally, we note that we have confined ourselves to the range
$\lambda\in(0,1]$. It is clearly possible to extend our study to
larger values of $\lambda$, in particular to
$\lambda\rightarrow\infty$, where the model converges to the Ising
model, but not without facing additional challenges, as follows. The
thermal fluctuations decrease with increasing $\lambda$, and the
effective interactions among the fields become weaker
\cite{kaupuzes2}. For large $\lambda$, the self-energy term of the
fields in the Hamiltonian becomes large. The step
  size has to be chosen small, otherwise it will lead to many rejected moves. As a consequence, the system gets
  trapped within narrow bands on the energy landscape.
  Our preliminary
  attempts to simulate the model at large $\lambda$ reveal that these
  traps give rise to artifacts (e.g., in force autocorrelation function at fixed $\Phi$) that are not easy to get rid
  of. These are issues we will explore in the future.

\section*{Acknowlegement \label{sec5}}

W.Z. acknowledges financial support from the China Scholarship Council (CSC).

\end{document}